\documentclass[amsmath,amssymb,aps,twocolumn,prb,showpacs]{revtex4}
\usepackage{graphicx}
\usepackage{dcolumn}
\usepackage{bm}
\usepackage{color}

\def\simge{\lower0.7ex\hbox{$\ \overset{>}{\sim}\ $}}
\def\simle{\lower0.7ex\hbox{$\ \overset{<}{\sim}\ $}}

\def\simge{\lower0.7ex\hbox{$\ \overset{>}{\sim}\ $}}
\def\simle{\lower0.7ex\hbox{$\ \overset{<}{\sim}\ $}}

\begin{document}

\title{p-Wave Superconductivity near a transverse saturation field}

%

\author{K. Hattori}
\author{H. Tsunetsugu}%
\affiliation{%
Institute for Solid State Physics, University of Tokyo, Kashiwanoha 5-1-5, Kashiwa, Chiba 277-8581, Japan
}%

\date{\today}

\begin{abstract}
We investigate reentrant superconductivity in an Ising ferromagnetic
 superconductor URhGe under a transverse magnetic
field $h_x$. 
The  superconducting transition temperature for $p$-wave order parameters $T_{sc}$ is
calculated and shows two domes as a function of
 $h_x$.
 We find strong
enhancement of $T_{sc}$ in the high-field dome near a saturation field $h_s$ where the spins align in
 the transverse direction. Soft magnons generate strong attractive
 interactions there. Spin components of the pairing show a significant
 change between $h_x$$<$$h_s$ and $h_x$$>$$h_s$. We also discuss the
 appearance of superconductivity with zero-spin pair due to cancellation between external
 and exchange fields.
\end{abstract}

\maketitle

Ferromagnetic superconductivity (SC) has attracted much
attention in condensed matter physics in the last decade.\cite{AokiFlouquet} U-based heavy-fermion compounds
such as UGe$_2$, URhGe, UIr, and UCoGe show unconventional
SC within their ferromagnetic phases.\cite{UGe2,Aoki,UIr,UCoGe} 
Non-unitary SC's\cite{Machida} are believed to appear in these
 compounds and both the SC and  ferromagnetism are caused in the
$f$ electrons at U sites. Their pairing mechanism and symmetry as well as 
 novel self-induced vortex states are central
issues to be clarified in the modern theory of unconventional superconductors.

Two isomorphic compounds, U{\it T}Ge({\it T}=Rh, Co), exhibit SC at
ambient pressure within their ferromagnetic state and have  
a similar Ising type anisotropy of magnetization.\cite{AokiFlouquet,Knafo} Spontaneous moment
appears parallel to the $c$-axis and its magnetization curve exhibits 
meta-magnetism when magnetic field ${\bf H}$ is applied to one of the hard
axes ($b$ axis) with a notable mass enhancement.\cite{Miyake} 
This meta-magnetism is particularly prominent in URhGe
and the moment 
gradually tilts with field and finally aligns parallel to ${\bf H}$  at $12$
T.\cite{LevyHuxley} Superconductivity appears below the transition temperature $T_{sc}$$=$$0.24$ K for ${\bf
H}$$=$${\bf 0}$, and $T_{sc}$ decreases with ${\bf H}$ and
disappears at $2$ T for $\bf H$ $\parallel b$.\cite{LevyHuxley} 
 Interestingly, SC  {\it reappears} above $8$ T
 and shows the highest $T_{sc}$$=$$0.42$ K at
 $12$ T.\cite{LevyHuxley}

Theories of ferromagnetic SC
have been developed by various
groups,\cite{Fay,Kirk,Mineev1,Mineev2,Tada,KlemmArXiv}
especially for UGe$_2$.\cite{Sandman,Nev,Linder,Uz}
In this paper, we focus on the reentrant SC in
URhGe, which has not been investigated theoretically. The physics of the
reentrant SC is quite different from
that in UGe$_2$.\cite{Sandman,Nev,Linder} 
 The point is the presence of {\it soft} magnons in the
 Ising systems with transverse fields. 
We clarify it by extending the Scharnberg-Klemm (SK)
theory\cite{SK}  in the presence of ferromagnetism and transverse
magnetic fields. In particular, we demonstrate (i) strong
enhancement of $T_{sc}$ near the saturation field, and (ii) strong field
dependence in spin components ($d$-vector) of $p$-wave pairing.

\begin{figure}[t]
\vspace{-1mm}
\begin{center}
    \includegraphics[width=0.5\textwidth]{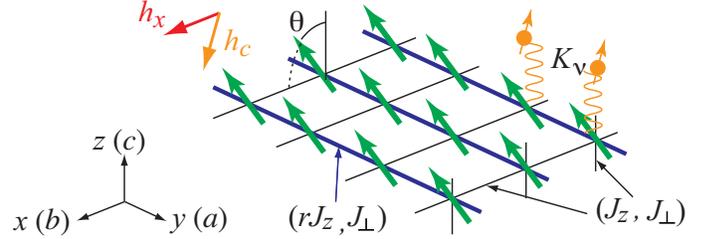}
\end{center}
\vspace{-3mm}
\caption{(Color online) Schematic picture of the model.}
\label{fig-1}
\vspace{-5mm}
\end{figure}

We consider a system of conduction electrons as shown in
Fig. \ref{fig-1}, coupled to ferromagnetic localized spins, which have 
a magnetization ${\bf M}$ in the $z$-direction 
(corresponding
to the $c$ axis in URhGe at ${\bf H}$$=$${\bf 0}$). We apply a transverse magnetic field $H_x$ 
along the $b$ axis and study its effects on ferromagnetic SC.
For conduction electrons, we use isotropic dispersion $\varepsilon_{\bf p}$$=$${\bf p}^2/2m^*$, 
ignoring complex band structures.\cite{bandcalc} Here, ${\bf p}$ and $m^*$ are
the momentum and the effective mass, respectively.

To model spin fluctuations in URhGe, we employ a simple model
that can 
describe their essential behavior, and that is 
the ferromagnetic XXZ model.\cite{single}  
To discuss SC, we will later introduce couplings between localized spins and
 conduction electrons. We treat local and itinerant degrees of
 freedom separately and this is qualitatively justified in the ``duality'' picture.\cite{Kuramoto}
There, ordered moments are due to the incoherent part of $f$ electrons,
 while itinerant physics such as SC is described by 
quasiparticles with heavy effective mass $m^*$. In this paper, we denote
magnetic moments and indices of the Kramers pairs simply as ``spin,''
but remember that these are complex combinations of orbital and spin moments in heavy-fermion systems.

The ferromagnetic XXZ model for localized moments is given as
\begin{eqnarray}
H_S\!\!\!&=&\!\!\! -\sum_{\langle i,j\rangle} \Big[J^z_{ij}S_{i}^zS_{j}^z
          +J_{\perp} \sum_{\nu=x,y}
	  S_{i}^{\nu}S_{j}^{\nu}\Big]
          -Sh_x\sum_iS_i^x,\ \ \ \  \label{XXZ}
\end{eqnarray}
where the spins ${\bf S}_i$ with ${\bf S}_i^2$$=$$S(S+1)$ are defined on a
three-dimensional cubic lattice and 
$h_x$$\equiv$$\mu H_x/S$. $\mu$$=$$0.4\mu_B$ is the value of magnetic
moment measured in experiment,\cite{Aoki} 
where $\mu_B$ is the Bohr magneton. 
We have introduced 
spatial anisotropy $J_{ij}^z$ as well as Ising spin anisotropy
$J_{ij}^z$$>$$J_{\perp}$$>$$0$ as shown in Fig. \ref{fig-1}, since the crystal
structure can be regarded as coupled-chains 
running in the $y(a)$ direction: $J_{ij}^z$$=$$J_z$ for bonds along $\hat{x}$ or
$\hat{z}$,  while $r J_z$ along $\hat{y}$ with
$r>0$. 
This spatial anisotropy 
lifts degeneracy in
$p$-wave superconducting order parameters as will be discussed
later.  Although the tetragonal symmetry of the exchange couplings is higher than 
 the orthorhombic one in URhGe, this is not essential for our discussions.

We analyze the model (\ref{XXZ}) for zero temperature $T$$=$$0$ by the
linearized spin-wave
approximation.\cite{SWA} Due to the Ising anisotropy, the spontaneous moment points along $\hat{z}$ at $h_x$$=$$0$.
 As $h_x$ increases, the  magnetization 
tilts toward $\hat{x}$ with polar angle
$\theta$$=$$\sin^{-1}(h_x/h_s)$  up to the saturation field $h_s$$=$$(4+2r)J_z-6J_{\perp}$. Magnons defined below describe transverse
fluctuations around this tilted direction.

The magnon energy dispersion is expressed as
$
E_{\bf q}$$=$$S\sqrt{{\epsilon_{\bf q}^2-4v_{\bf q}^2}},
$
where 
%
%
%
%
$\epsilon_{\bf q}$$=$$(c_{\theta}^2$$-$$s_{\theta}^2/2)h_s$$+$$h_xs_{\theta}$$-$$J_{\perp}(c_{\theta}^2$
$+$$1)\gamma_{\bf q}$$-$$J_zs_{\theta}^2\gamma_{\bf q}^y$,
$
2v_{\bf q}$$=$$s_{\theta}^2(h_s/2$$+$$J_z\gamma_{\bf q}^y$$-$$J_{\perp}\gamma_{\bf q})$ 
with 
$\gamma_{\bf q}$$=$$\sum_{l=x,y,z}$$(\cos q_{l}$$-$$1)$, $\gamma_{\bf
q}^y$$=$$\gamma_{\bf q}$$+$$(r$$-$$1)$$(\cos q_y$$-$$1)$, and $c_{\theta}(s_{\theta})$$=$$\cos\theta(\sin\theta)$.
%
The dispersion has an energy gap   
$S\sqrt{h_s^2-h^2_x}$ for $h_x$$<$$h_s$ and 
$S\sqrt{h_x(h_x-h_s)}$ for $h_x$$>$$h_s$. At $h_x$$=$$h_s$,    
 the excitations are gapless 
$E_{\bf
q\sim 0}$$\sim$$S\sqrt{h_s
J_z(q_x^2+r q_y^2+q_z^2)}$.   
 Note that the model (\ref{XXZ}) is
ferromagnetic, and thus the linearized spin-wave approximation is
fairly good   
even near
$h_s$. 
Since the magnitude of $S$ does not play an important role in our
approximation, we set $S$$=$$1$.
The renormalization of moment due to the quantum
fluctuation is small and we neglect it hereafter.

%

Let us investigate the interactions of magnons and conduction
electrons (quasiparticles). 
The quasiparticles 
feel the external field and exchange field via anisotropic
antiferromagnetic Kondo coupling 
$K_{x,y,z}$: $\sum_{i\nu} K_{\nu}s_i^{\nu}S_i^{\nu}$ with
$s_i^{\nu}$ being the spin of the quasiparticle at the site $i$. 
Again, note that this ``spin'' is, indeed, ``pseudospin''.
Thus, in the mean field level, their energy
dispersion is $\varepsilon_{\bf p\pm}$$=$$\varepsilon_{\bf p}\pm
|{\bf h}_c|/2$ with 
$
{\bf
h}_c$$=$$(h_c^x,0,h_c^z)$$\equiv$$(\mu_cH_x$$-$$K_{x}s_{\theta},$$0,-K_zc_{\theta})$.
Here, $\mu_c$$\equiv$$g^*\mu_B$ with the effective $g$-factor of
quasiparticles $g^*$.

The lowest-order fluctuations beyond the mean-field approximation yield 
a magnon-quasiparticle interaction:
$\sum_{i\nu}$$v_{\nu}a_i^{\dagger}{\tilde s}_i^{\nu}$+H.c.,
where $a^{\dagger}_i$ is a magnon creation operator at the site $i$ and 
$\tilde{s}_i^{\nu}$ is the quasiparticle spin rotated such that the new
$z$-axis is parallel to $-{\bf h}_c$. 
The coupling constant $v_{\nu}$
is given as 
$(v_x,v_y,v_z)$$=$$(K_zs_{\theta}\hat{h}_c^x$$+$$K_xc_{\theta}\hat{h}_c^z$,$-iK_y,$$K_zs_{\theta}\hat{h}_c^z$$-$$K_xc_{\theta}\hat{h}_c^x)/\sqrt{2}$
with $\hat{h}_c^{x,z}$$\equiv$$h^{x,z}_c/|{\bf h}_c|$.


Next, we derive linearized gap equations for $p$-wave
SC\cite{SK} and
 determine the transition temperature $T_{sc}$ and 
upper critical field $H_{c2}$. Since we are mainly interested in
high-field states, we neglect small $4\pi {\bf M}$ less than $\sim$ 1 kG in the
magnetic induction.\cite{Mineev1} This must be taken
into account near $h_x$$=$$0$ and is important for discussions about
self-induced vortex states, which is one of our future problems.
Since ${\bf H}\parallel \hat{x}$, the gap equation for the $p_x$ state
 $\vec{\Delta}_{xN}$$=$$[\Delta_{xN}^{++},\Delta_{xN}^{--},\Delta_{xN}^0]^T$
 and the other orbitals
$\{\vec{\Delta}_{yN},\vec{\Delta}_{zN}\}$ are
decoupled, where $N$ represents the Landau level (LL), and  ``$++$'',
``$--$'', and ``$0$'' denote the spin projection of Cooper
pairs along $-{\bf h}_c$. Different LL's for the $p_x$ state are
decoupled to each other, while
they are coupled for the $(p_y, p_z)$ components.
In the SK theory, the linearized gap equations  are given as\cite{SK}
\begin{eqnarray}
\vec{\Delta}_{\alpha N}=\frac{3T{m^*}^2}{2\pi}\hat{V}_{\alpha}\sum_{N',
 \beta}\hat{S}_{NN'}^{\alpha\beta}\vec{\Delta}_{\beta N'},\quad
 (\alpha,\beta\in \{x,y,z\}). \label{GapEq}
\end{eqnarray}
Here, $\hat{V}_{\alpha}$ is an interaction matrix of the spin part [see
 Eq.(\ref{tripletV})] for the $p_{\alpha}$ state.  
 $\hat{S}$ 
includes both the Zeeman and orbital pair breaking
 effects and is diagonal in spin space.


$\hat{V}_{\alpha}$ is determined from the one-magnon exchange processes
and we calculate this with the static approximation for the magnon
Green's function 
$D_{\bf q}(i\omega_m)$$=$$1/(i\omega_m-E_{\bf q})$$\sim$$-1/E_{\bf q}$. 
Averaging all the momentum dependence
assuming $p$-wave gap functions $\eta_{\bf p}^{\alpha}$, $\hat{V}_{\alpha}$ is given as
%
\begin{eqnarray}
\hat{V}_{\alpha}\!\!\!&=&\!\!\!
A_{\alpha}^+v_z^2\!
\begin{pmatrix}
1&
w^2&
\sqrt{2}w\\
w^2 &1 & -\sqrt{2}w\\
\sqrt{2}w&-\sqrt{2}w&
w^2-1\\
\end{pmatrix}
+
A_{\alpha}^-v_y^2
\begin{pmatrix}
0&
1&
0\\
1 &0 & 0\\
0&0&
-1
\end{pmatrix},\nonumber\\
&\equiv& A_{\alpha}^+v_z^2 \hat{V}_+(w)+A_{\alpha}^-v_y^2 \hat{V}_-, \label{tripletV} 
\end{eqnarray}
where $w$$=$$v_x/v_z$ and the first and the second columns of $\hat{V}_{\alpha}$
represent the equal spin pairing (ESP)  
with the spin function $|$$++\rangle$ and $|$$--\rangle$, respectively, while the third one
represents the zero-spin pairing (ZSP) state $(|$$+-\rangle$$
+$$|$$-+\rangle)/\sqrt{2}$. Note that ``$-$'' and ``$+$'' denote spins parallel
and antiparallel to ${\bf h}_c$, respectively. 
Here, 
\begin{eqnarray}
A_{\alpha}^{\pm}&\equiv&\frac{1}{N_L^2}\sum_{\bf p,k}\eta_{\bf p}^{\alpha}\frac{e^{\pm
 2\phi_{\bf p-k}}}{E_{\bf
 p-k}}\eta_{\bf k}^{\alpha},
\end{eqnarray}
where the pairing form factor is chosen as $\eta_{\bf
 p}^{\alpha}$$=$$\sqrt{2}\sin p_{\alpha}$ and this corresponds
 to $\sqrt{3}p_{\alpha}$ in continuum systems.\cite{SK}
$\phi_{\bf p-k}$ is related to the Bogoliubov transformation: 
 $\tanh 2\phi_{\bf p-k}$$=$$2v_{\bf
 p-k}/\epsilon_{\bf p-k}$, and $N_L$ is the number of sites. 

Before showing numerical results, we analyze $\hat{V}_{\alpha}$ for $h_x\sim
h_s$. Since $E_{\bf q}$ is small around ${\bf q}$$=$${\bf 0}$,
$T_{sc}$ is expected to be enhanced. It is important to note that  
the direction of the
local moments continuously changes toward $\pi/2$ for $h_x<h_s$, while $\theta$$=$$\pi/2$
above $h_s$. Expanded  
up to first order in $\delta\theta$$=$$\pi/2-\theta$, $\hat{V}_{\alpha}$ is given  as
\begin{eqnarray}
\frac{1}{2}
\begin{pmatrix}
0&
A_{\alpha}^+K_z^2-A_{\alpha}^-K_{y}^2&
-\Gamma A_{\alpha}^+ \delta\theta\\
A_{\alpha}^+ K_z^2-A_{\alpha}^- K^2_{y} &0 & \Gamma A_{\alpha}^+ \delta\theta\\
-\Gamma A_{\alpha}^+ \delta\theta &\Gamma A_{\alpha}^+ \delta\theta &
 A_{\alpha}^+ K_z^2+{A_{\alpha}^-} K_{y}^2
\end{pmatrix}, \label{approx}
\end{eqnarray}
where
$
\Gamma$$=$$\sqrt{2}K_z^2\{
[\mu_c h_s/(\mu K_z)-K_x/K_z]^{-1} -K_x/K_z
\} 
$.
For $h_x$$\ge$$h_s$, 
 $\delta\theta$$=$$0$, and the
gap equations for ESP and ZSP are decoupled. 
  Note that the interaction is strongest for ZSP,
 since $A_{\alpha}^{\pm}$$>$$0$. This is natural, since the spin fluctuations
 are transverse, which scatter quasiparticles with $\pm$ spin
  into $\mp$ spin states.

Low-field longitudinal fluctuations\cite{T-Hattori} also mediate
interactions 
and we introduce a phenomenological ferromagnetic Ising interaction between
quasiparticles: $-\sum_{\langle i,j\rangle}J_{cc}^{\alpha}s_i^zs_j^z$, 
with $J_{cc}^{\alpha}$$>$$0$ and $\alpha$$\in$$\{x,y,z\}$ is the direction of
 $ij$ bond. As before, we include spatial anisotropy: 
 $J^{ij}_{cc}$$=$$J_{cc}$ for bond $\parallel{\hat{x}}$ or
${\hat{z}}$, while $r'J_{cc}$ for bond $\parallel\hat{
y}$. Then, the interaction kernel (\ref{tripletV}) is replaced as
$\hat{V}_{\alpha}$$\to$
$\hat{V}_{\alpha}+({\hat{h}_z^c})^2 J_{cc}^{\alpha}\hat{V}_+({\hat{h}_x^c}/{\hat{h}}_z^c)/4$,
 with keeping the form (\ref{approx}) essentially unchanged.

\begin{figure}[t!]
\begin{center}
    \includegraphics[width=.5\textwidth]{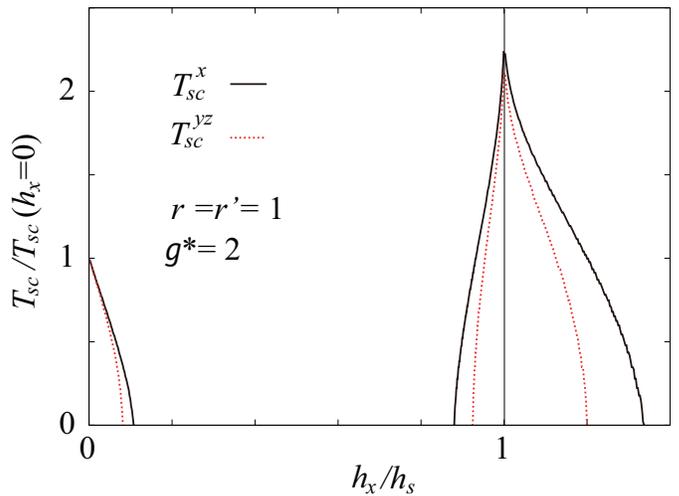}
\end{center}
\vspace{-3mm}
\caption{(Color online) Dependence of transition temperature 
  $T_{sc}^{x,yz}$  on transverse field $h_x$ 
for $r$$=$$r'$$=$$1$ and $g^*$$=$$2$.} 
\label{fig-2}
\vspace{-4mm}
\end{figure}

\begin{figure}[t!]
\begin{center}
    \includegraphics[width=.5\textwidth]{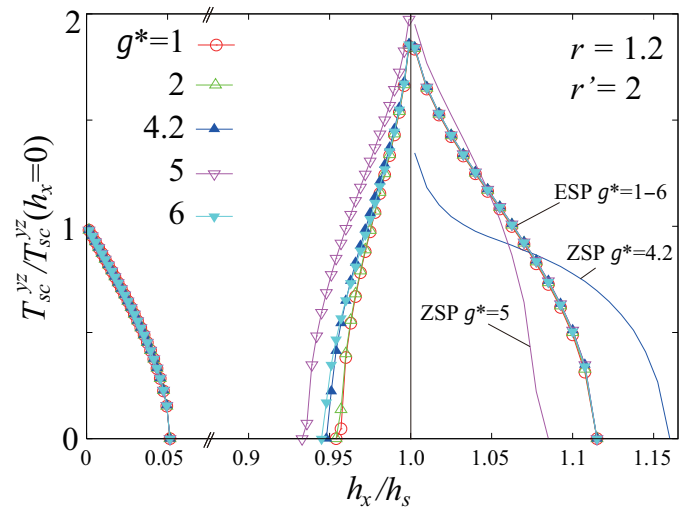}
\end{center}
\vspace{-3mm}
\caption{(Color online) $h_x$-dependence of  $T_{sc}^{yz}$ for the
 anisotropic case $r$$=$$1.2$,
 $r'$$=$$2$ and five sets of $g^*$. The change of low-field dome with
 $g^*$'s is negligible.
 For $h_x$$>$$h_s$, the ZSP state is
 decoupled from the ESP state. Its
$T^{yz}_{sc}$ is shown by lines without symbols, except for
 $g^*$$=$$1,2$, and 6, where $T^{yz}_{sc}$ is nearly zero.
}
\label{fig-3}
\vspace{-5mm}
\end{figure}

\begin{figure}[tb]
\begin{center}
    \includegraphics[width=.5\textwidth]{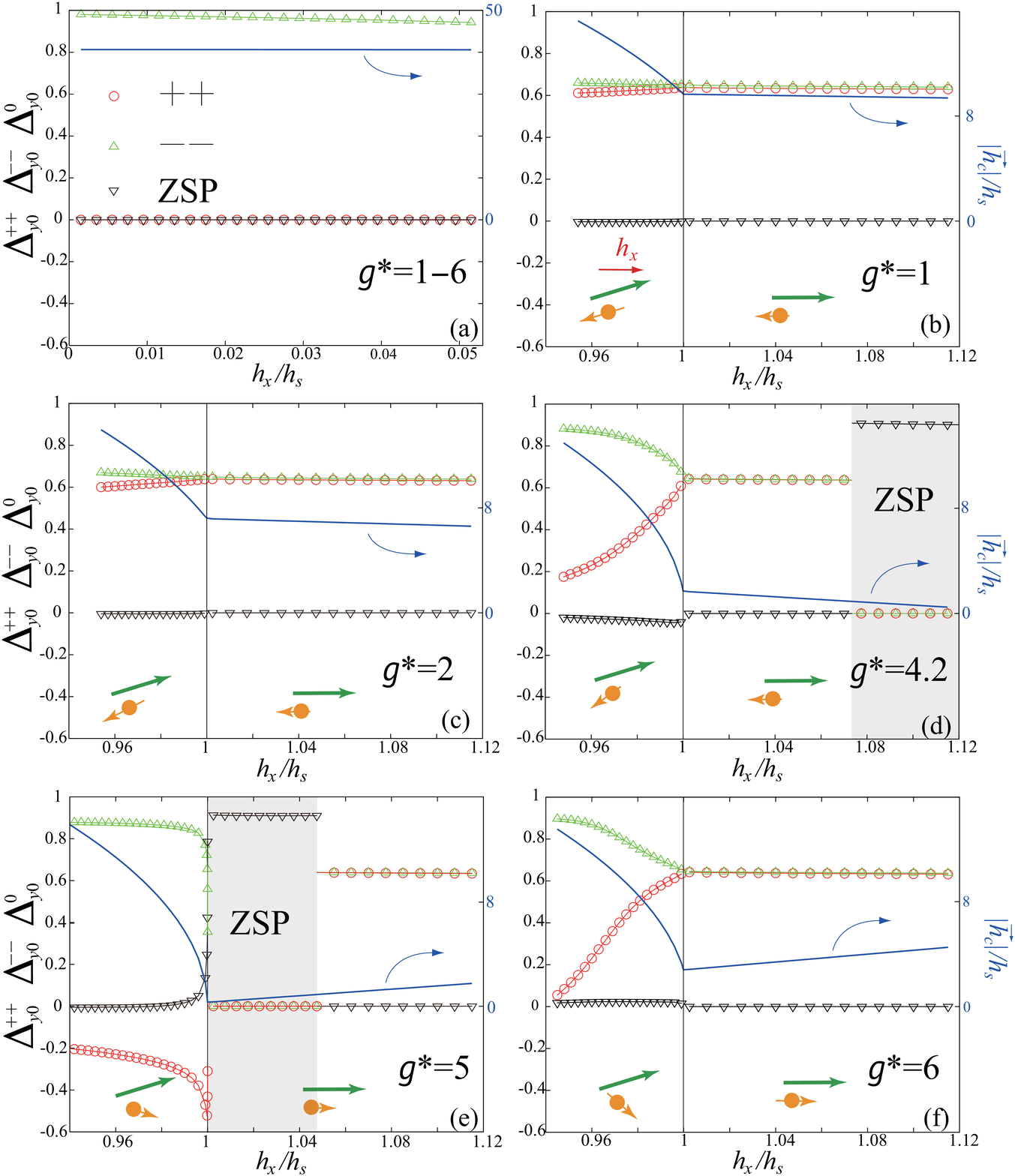}
\end{center}
\vspace{-3mm}
\caption{(Color online) Spin part of $p_y$
pairing in the $N$$=$$0$ LL for $r$$=$$1.2$ and $r'$$=$$2$:
 $\Delta_{y0}^{++}(\bigcirc)$, $\Delta_{y0}^{--}(\bigtriangleup)$, and
$\Delta_{y0}^{0}(\bigtriangledown)$ at $T$$=$$T_{sc}^{yz}$. 
Lines represent $|{\bf h}_c|$. 
 (a) Low-field part, $g^*$-dependence is negligible. 
(b)--(f) High-field part near $h_s$ for various $g^*$'s. 
Localized moment (long arrow) and quasiparticle spin (short one with
 dot) are schematically shown. The magnitude of the
 quasiparticle spin is small $\le 0.05$ near $h_s$.}
\label{fig-4}
\vspace{-3mm}
\end{figure}

Let us discuss the phase diagram of SC in the
$T$-$h_x$ space determined by our calculations. Figures \ref{fig-2} and
\ref{fig-3} show the $h_x$-dependence of the SC transition temperature
of the $p_{x}$- and $p_{y,z}$-states for the isotropic $(r$$=$$r'$$=$$1)$ and
anisotropic $(r$$=$$1.2,r'$$=$$2)$ cases, respectively. Figure \ref{fig-4} shows
the spin part of the pairing, and will be discussed later. 
 Our calculations show two domes of $T_{sc}$ in the phase
diagram, and this qualitatively reproduces the experimental data.\cite{LevyHuxley} One
important observation is a strong enhancement and the presence of
singularity in the high field dome at $h_x$$=$$h_s$. This manifests 
 the change in the pairing state. The variations in the spin
components are due to competition between the interaction strength and
the Pauli depairing effects (PDE's) and we will examine this point in
detail in the following.

Let us investigate how the PDE's influence pairing states. 
The PDE in the SK theory originates in the matrix $\hat{S}$ in
Eq.(\ref{GapEq}).  
For $h_x$$>$$h_s$, the ESP's and ZSP are decoupled as discussed above,
which simplifies analysis. 
The PDE is weak when the exchange field
nearly cancels the external field $h_c^x \sim 0$. This is realized at high field, and
we expect there the ZSP state, since the corresponding interactions are
 strongest. This is the Jaccarino-Peter effect,\cite{JPeffect} which was originally
 proposed for a rare earth ferromagnetic metal. When the cancellation is
 not sufficient and thus PDE is strong,
 the ZSP state is suppressed and the ESP state is favored. If the
 interactions are much stronger than the effective magnetic field, two
 ESP components $|$$++\rangle$ and $|$$--\rangle$, have nearly equal
 amplitude. If the interactions are not so strong, the ESP state with
 larger density of states (DOS) dominates

For $h_x\simle h_s$, the situation is more complex.  
 When PDE is strong, the ESP with large DOS is realized as expected also
 for low field, while for weak PDE all the spin
components contribute to the superconducting condensation, since the
 offdiagonal elements in $\hat{V}_{\alpha}$ are finite.

Now, we explain the details of Figs. \ref{fig-2}--\ref{fig-4}. Our choice of parameters 
are: a mean-field ferromagnetic transition temperature $T_c^{\rm
 MF}$$=$$J_z(2+r)$$=$$10$ K, $H_s$$\equiv$$h_s/\mu$$=$$10$ T, $m^*/m$$=$$40$,
 $r'J_{cc}$$=$$283$ K,
$K_z$$=$$109$ K, and $K_{x,y}$$=$$0.3K_z$, where $m$ is the bare electron
mass. Electron density is set to 1 per unit cell and the lattice
constant\cite{Aoki} is set to $3.5$ \AA.
Using these parameters, $T_{sc}$ at $h_x$$=$$0$ is $\simeq$$0.6$ K for both
parameter sets in Figs. \ref{fig-2} and \ref{fig-3}.\cite{logOm}   $g^*$ is the 
control parameter in our analysis below, which varies the Zeeman energy
for quasiparticles.

Let us first examine the orbital part of pairing. 
The orbital
motion couples only to the external field $H_x$ as we mentioned above. 
Therefore, the
$p_x$-pairing is always decoupled from the other orbitals and different
LL's are decoupled there, while the $p_y$ and $p_z$
orbitals are mutually coupled with LL's also coupled. In the isotropic
systems, the polar $p_x$-state with $N$$=$$0$ has the highest $H_{c2}$ as
shown in Fig. \ref{fig-2}.
We note
that the $N$$=$$0$ LL plays a central role in both $p_x$ and $p_{y,z}$ states.\cite{SK}



When the interactions are spatially anisotropic in space with $r, r'$$\ge$$1$,
the $p_x$ state is suppressed and a $p_y$-dominant
state becomes stable, and {\it vice versa} for $r,r'$$\le$$1$. In the
following, we analyze the former case, which is relevant for URhGe.
Figure \ref{fig-3} shows $T_{sc}^{yz}$ for
$r$$=$$1.2$ and $r'$$=$$2$, as a typical case of strong anisotropies, for several values of
$g^*$. Here, $T_{sc}^x\ll T_{sc}^{yz}$ for all the parameters and not shown.  
For the low-field SC, a $p_y$ dominant state has a lower energy even for small
$r'$$\simge$$1.05$, while less sensitive to $r$. This is because 
that the magnon gap is sufficiently large there and thus the
transverse fluctuations are negligible.
For the high-field SC, both $r$ and $r'$ affect its $T_{sc}$. Increasing 
$r$ enhances $T_{sc}^{yz}$, while it suppresses $T_{sc}^x$. For the
dependence on 
 $r'$, increasing $r'$ suppresses both $T_{sc}$'s. 
The suppression is stronger for $T^x_{sc}$ than for $T_{sc}^{yz}$ because the
attractive force does not decrease for $p_y$ pairs. It is also possible
to realize a transition in the SC from $p_x$ to $p_{y,z}$ symmetry or {\it vice
versa}, as $h_x$ increases, which is determined by the details of the anisotropies.

Second, we discuss the spin part of pairing, which is related to
the $d$-vector of SC, and examine for the data in Fig. \ref{fig-3} 
the Zeeman effects with controlling $g^*$.
 Figure \ref{fig-4} shows the change in the spin part of pairing with
 $h_x$ for the orbital component $p_y$ and $N$$=$$0$ LL. They are calculated
 just at $T_{sc}^{yz}$ and 
 normalized as $\sum_N$$\sum_{\alpha}$$|\vec{\Delta}_{\alpha N}|^2$$=$$1$.
Figure \ref{fig-4}(a) shows the change in the low-field dome. 
We find that the $g^*$-dependence is negligible and the pairing is the ESP state with
only one type of spin, which is parallel to the internal field ${\bf
h}_c$, $|$$--\rangle$. This is consistent with
 the proposed
 gap symmetry for URhGe.\cite{HardyHuxley} 
Figures \ref{fig-4}(b)--\ref{fig-4}(f) show the results in the high-field dome. 
The region of $h_x$$>$$h_s$ is, in most of the cases, also the pure ESP
 state, but now the two spin components have almost the same amplitude. 
This is because $|$$++\rangle$ pairs are scattered only to
 $|$$--\rangle$ pairs and {\it vice versa}, when $h_x$$>$$h_s$. However,
 their amplitude difference may become larger depending on $g^*$ and
 $|{\bf h}_c|$. In the region of $h_x$$<$$h_s$, a small amplitude of the
 ZSP
 component hybridizes. The two ESP components are still dominant, but
 their difference is now larger and the $h_x$-dependence is noticeable,
 particularly for larger $g^*$'s. 
%
For all $g^*$'s, transverse spin fluctuations near $h_s$ favor the linear
 combinations of two ESP states, while ZSP is suppressed by the PDE, except for
 some regions in Figs. \ref{fig-4}(d) and \ref{fig-4}(e) as we 
 discuss below. 


The quasiparticle spins respond to both the exchange field of the localized
moments and the external magnetic field, and $g^*$ controls the latter part.
Thus, varying $g^*$ affects the angle between the quasiparticle 
and the localized spins as schematically depicted in
Figs. \ref{fig-4}(b)--\ref{fig-4}(f). When the external and exchange fields are
canceled, ${\bf h}_c$$=$${\bf 0}$, the quasiparticle spins feel no field. 
Indeed, the cancellation can become nearly perfect for $g^*$$=$$4.2$
 and $5$. 
This leads to emergence of ZSP states
(Jaccarino-Peter effect),\cite{JPeffect} and 
 occurs in the shaded regions in Figs. \ref{fig-4}(d) and
\ref{fig-4}(e). 
When $h_s$ is close to the point $|{\bf h}_c|$$\sim$$0$
 as in Fig. \ref{fig-4}(e),  the expansion in 
Eq.(\ref{approx}) is not sufficient, and this results in the almost
vertical slope of $\vec{\Delta}_{y0}$ for $h_x$$\simle$$h_s$, but
this situation is not of main interest in this paper.

Let us finally comment about URhGe. We have succeeded in reproducing
the overall phase diagram and two SC domes as shown in Figs. \ref{fig-2} and
\ref{fig-3}. The transition temperature for the second dome
shows strong enhancement near $h_s$. The superconducting gap symmetry is
basically equal-spin $p_{y(a)}$-pairing for both domes. It is $|$$--\rangle$
for the low-field, and $\sim
($$|$$++\rangle$$+$$|$$--\rangle$$)/\sqrt{2}$ for the high-field SC.
The gap functions have a point node due to the finite $ip_z$
component, but since its amplitude is very small, 
 it might be difficult to experimentally distinguish
this from that with a line node. 
Detecting the change in the spin components of pairing  
near $h_s$ is an experimental test for the present
theory. 
Exploring the sudden change in pairing symmetry due to the
Jaccarino-Peter effect is another interesting challenge in this field.

For a more precise theoretical analysis, one should take into account the
optimization of form factor $\eta_{\bf k}^{\alpha}$, longitudinal spin 
fluctuations for $h_x\simle h_s$,\cite{LevyHuxley} the weak first-order
transition,\cite{LevyHuxley} the variations of the Fermi
surfaces near $h_s$,\cite{Lif} and anisotropy in the effective mass,\cite{KlemmArXiv} 
which 
might improve quantitative agreement between our theory and experiments.
Our main conclusion for the
superconducting order parameters for $h_x$$\sim$$h_s$ is that the pairing
interaction prefers mixing $|$$++\rangle$ and $|$$--\rangle$ and this is 
in clear contrast to nearly pure $|$$--\rangle$ state, which is selected
by Zeeman energy.
Difference in their amplitudes depends on the details, 
and a more quantitative analysis about this is one of our future plans.



K. H. thanks Dai Aoki for fruitful discussions.
He is supported by KAKENHI (No.\! 30456199) and by a Grant-in-Aid for Scientific Research on Innovative
Areas ``Heavy Electrons'' (No.\! 23102707) of MEXT, Japan.


\begin{thebibliography}{99}
\bibitem{single} Essentially the same results are obtained in 
	presence of single-ion anisotropy.
\end{thebibliography}
\end{document}